# PREHISTORIC SANCTUARIES IN DAUNIA[1]

ELIO ANTONELLO[1], VITO F. POLCARO[2], ANNA M. TUNZI[3], MARIANGELA LO ZUPONE[3]

[1]INAF-Osservatorio Astronomico di Brera;     e-mail:  elio.antonello@brera.inaf.it
[2]INAF-IASF, Rome
[3]Soprintendenza per i Beni Archeologici della Puglia

**Abstract:** Daunia is a region in northern Apulia with many interesting sites particularly of the Neolithic and Bronze Age. Beginning from the fifth millennium BC, the farmers living in the wide plain of Daunia dug hypogea and holes in possible prehistoric sanctuaries. The most impressive phenomenon is the presence of rows of hundreds holes dug in the stratum of calcareous rock located under few tens centimetres of soil. The characteristics of the holes suggest a ritual use, and the archaeologists tend to exclude other applications such as cultivations and post holes. The archaeoastronomical analysis of the first sanctuary discovered in Trinitapoli indicated few specific astronomical directions for the rows. In 2009 a new sanctuary with some Neolithic hypogea was discovered in Ordona. Just two small areas were excavated, and no remains were found inside the hole rows, so no estimate of their probable age is possible. The extent of the sanctuary is not known, but it should cover a very large area as in the case of Trinitapoli, suggesting the use of such a site for many centuries. Differently from the essentially solar and lunar criteria possibly adopted for the alignments in Trinitapoli, the community of Ordona possibly used some stars of the Centaurus – Crux group (may be α Centauri itself), and we cannot exclude the presence of the effects of the precession on the stellar alignments of the rows. In the past centuries, several astronomers and scholars remarked the spectacular region of the sky of Centaurus – Crux group; its possible relevance for ancient civilizations was pointed out by G.V. Schiaparelli in his work on the astronomy in the Old Testament, where he mentioned in particular the observations of the astronomer W. S. Jacob. It would be worth to make simulations in order to reproduce the 'Jacob effect', that is the diffused light produced by that rich stellar region of the sky.

## 1. Introduction

In the Italian region of Apulia (Puglia) there are many archaeological sites dating from Neolithic to Roman and Medieval epoch, and some finds date back to the Palaeolithic period. Beginning from the 5th millennium BC and for more than three thousand years the farmers living in the wide plain of the Daunia region in northern Apulia dug hypogea and holes for ritual and funerary purposes. Up to now two sanctuaries have been found in Trinitapoli and Ordona. In the Bronze Age sanctuary of Trinitapoli there are some hypogea, but the most impressive phenomenon is the presence of rows with hundreds holes dug in the stratum of calcareous rock located below few tens centimetres of soil. The characteristics of the holes and the remains found inside indicate a ritual use, while other applications such as post holes and cultivations should be excluded. The astronomical analysis suggests that the ancestors may have adopted few specific astronomical directions for the alignments of the rows, and these results were discussed at the SEAC 2008 meeting (Tunzi et al., 2009; for further details, see also Tunzi et al., 2010). The area interested by the phenomenon is very large, as shown by the most recent excavations. An obvious suspicion is that the holes were dug for tree plantation, e.g. vines as described by the Latin writer Columella (*De re rustica*, 1st century AD). However, several features such as the characteristics of the bedrock, the geometry of the holes, their short spacing, the age of the prehistoric remains found in some of them and the too early prehistoric epoch for systematic tree plantations should exclude such an interpretation.

---

[1] Paper presented at the SEAC (European Society for Astronomy in Culture) 2010 meeting in Gilching (Germany).

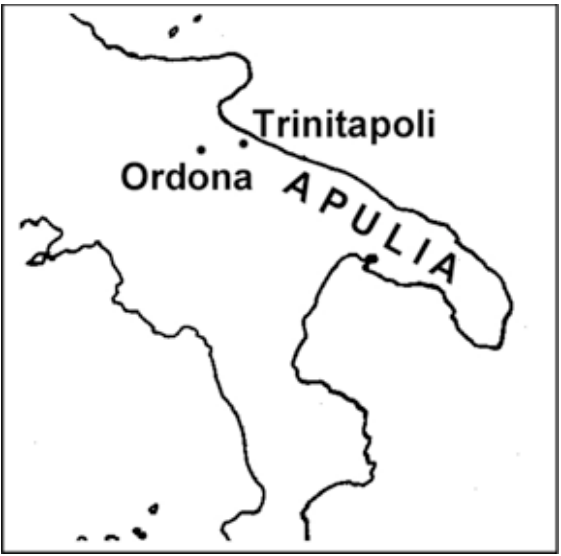

**Fig. 1.** Southern Italy and the location of the archaeological sites of northern Apulia discussed in the present work.

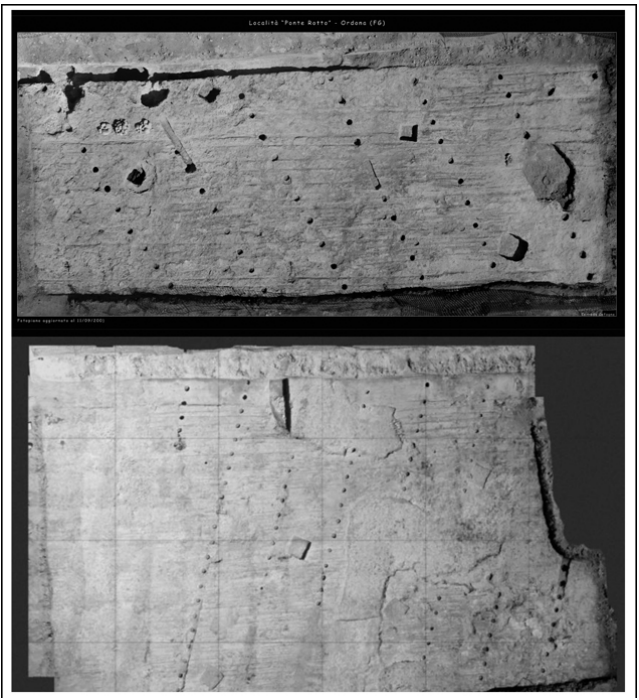

**Fig. 2.** Orthophoto of the two areas of Ponterotto 1 (upper panel) and Ponterotto 2 (lower panel) of Ordona. The holes appear as 'dots'. The separation between two adjacent holes is less than one meter, and the distance between the rows is of about three meters.

## 2. Ordona

Last year a new Neolithic sanctuary was discovered in Ordona, about thirty kilometres west of Trinitapoli. The occasion was the installation of a cable connecting the wind turbines of a wind farm. Only two small areas were excavated, named Ponterotto 1 and Ponterotto 2. Straight rows of holes dug quite carefully were found in both areas. The separation between the circular holes (with a diameter of about thirty centimetres) is generally less than one meter, and that between the rows more than three meters. The largest side of the two excavated areas is of the order of twenty meters; the two areas are located rather far apart, about nine hundred meters. The remains found in some hypogea were dated at the mid 5th Millennium BC. The holes however contained just soil, and for the present there is no reliable estimate of their age. The real extent of the sanctuary is not known, however it may cover a very large area as in the case of Trinitapoli, and it may be possible it was similarly used for many centuries.

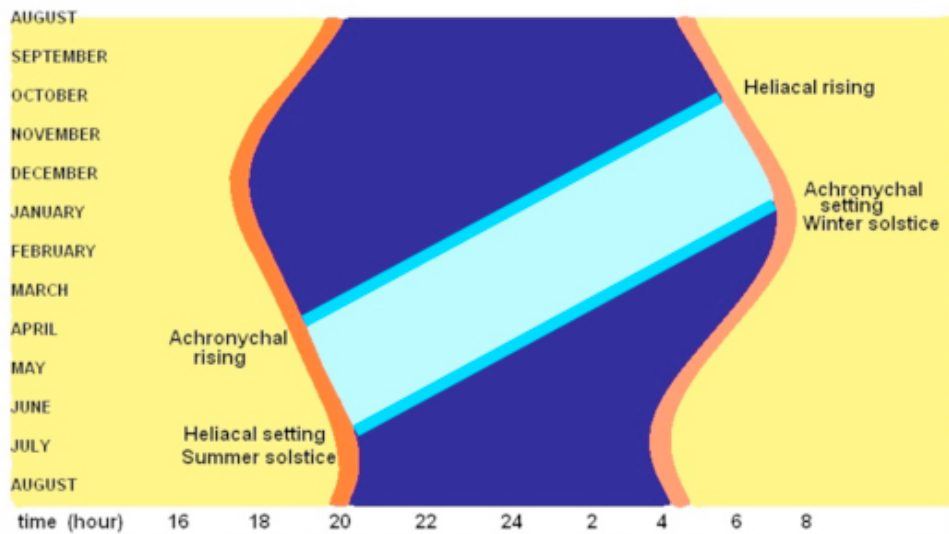

**Fig. 3.** Qualitative plot of the visibility of the Centaurus-Crux stars during the Neolithic – Bronze Age in Southern Italy. The visibility is indicated by the band in light color, while the darkest color indicates the night time (from evening to morning) along the year (the month is shown on the left).

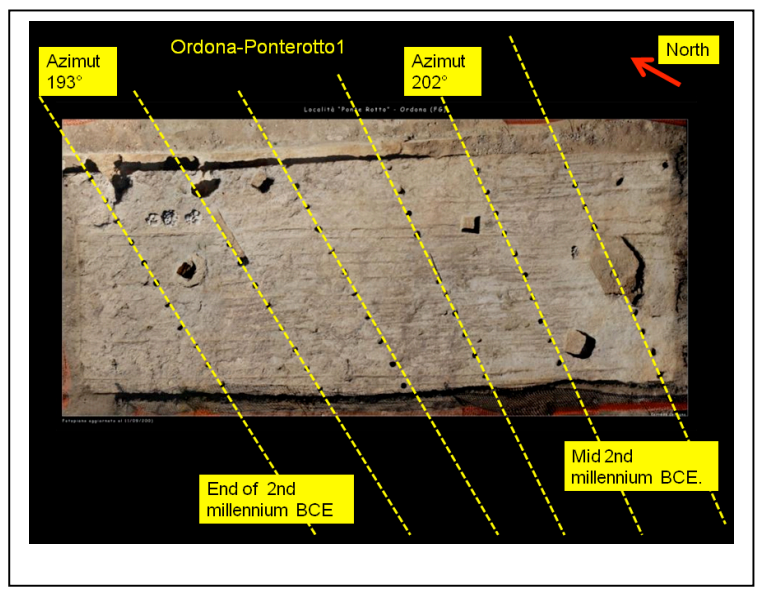

**Fig. 4.** Orientation and hypothetical dates of the hole rows of the site of Ponterotto 1 (Ordona).

The orientation of the rows is between 194 and 213 degrees in South-West direction (or between 14 and 33 degrees in the North-East direction). Here we present a preliminary analysis. The azimuth values suggest that, differently from the essentially solar and lunar criteria possibly adopted for the alignments in Trinitapoli, the community of Ordona may have used the setting of some specific stars. We considered the South-West direction, since the astronomical phenomena in the North-East direction do not appear to be adequate to explain the observed azimuths. We suspect that the southern sky stars belong to the group of Centaurus – Crux. At the time of the hypogea, the stars of this group set behind the nearby mountains of Deliceto, while during the 3rd millennium BC the direction of the setting could be that indicated by the rows of Ponterotto 2. The rows of the other area, Ponterotto 1, could indicate the setting of the same stars occurred a millennium later. The group of stars was visible in Southern Italy, during Neolithic – Bronze Age, beginning from October till the end of spring (present day calendar, Figure 3). For example, at mid October there was the heliacal rising of γ Crucis while that of α Centauri occurred some days later. The achronychal setting of the stars (just before the sunrise) occurred at the time of the winter solstice, and the heliacal setting occurred some days before the summer solstice. One should note that this group of stars was visible during wintertime. We cannot exclude for the present the effect of the precession on the alignments, which could explain indeed the progressively slightly different directions of the rows in the area of Ponterotto 1. Assuming α Centauri as the target star of the ancestors, the holes of Ponterotto 2 should have been dug at the mid of the 3rd millennium BC, and those of Ponterotto 1 from the mid to the end of the 2nd millennium BC (Figure 4). Of course, this is just a preliminary analysis, and further archaeological discoveries could change this scenario; in particular, an archaeological dating is needed before drawing any reliable conclusion.

## 3. The Centaurus – Crux region

As a support to the Centaurus-Cross interpretation, we recall that Hoskin (2001, 42-51) pointed out the possible importance of these constellations for the sanctuaries of the Mediterranean basin. He tacitly assumed, as we did, that the ancestors were just interested in the bright stars of the constellations. This looks rather obvious, but this is an assumption. More than one hundred years ago, Schiaparelli published an interesting study on the astronomy in the Old Testament, which was quickly translated in German and English (Schiaparelli 1903), and which contains a potentially important discussion regarding this issue. As we will try to show, it could put the issue literally "in a new light".

Schiaparelli discussed carefully the identifications of some constellations and asterisms in the Old Testament, according to the different ancient versions and translations of the Bible. A verse in the Book of Job (Job 9, 9) quotes the Bear (*hasc*), Orion (*kesil*) and (*we*) the Pleiades (*kimah*). Moreover, the last two words of the verse, *chadre teman*, were translated by Schiaparelli as chambers of the South. That is, the verse should be similar to that in the New Revised Standard Version 1989 of the Bible: *who made the Bear and Orion, the Pleiades and the chambers of the South*. Schiaparelli pointed out that the word *chadre* should indicate the innermost and private chambers of a building, while *teman* indicated both the right and the southern direction. Schiaparelli probably was exceedingly enthusiast when he declared that the author of the Book of Job unquestionably wished to indicate, by these two words, some brilliant southern constellation; actually, nobody should be able to know what the author of the Book of Job really wished (or wanted) to indicate. Schiaparelli discussed at length this spectacular case, though he never travelled beyond Europe, therefore he never saw the southern constellations. What is the reason for his enthusiasm? He described the richness of both bright and faint visible stars in a region of the southern sky from α Argus (Canopus) to α Centauri, taking into account the studies of the astronomers and scholars of his times. The star α Argus is today α Carinae, since Argus has been divided in Carina, Vela and Puppis. This part of the sky is *the splendour of the southern heavens*, as declared by von Humboldt (1858, 146-147). Schiaparelli moreover quoted his own study on the distribution of visible stars, and he wrote that this part of the sky produces in the atmosphere a sort of twilight illumination. Indeed he took from the book of von Humboldt an impressive description (reported in English even in the original Italian edition of his book) of the effect of the increase of diffused light when the Southern Cross was risen: *Such is the general blaze of star-light near the Cross, from that part of the sky, that a person is immediately made aware of its having risen above the horizon, though he should not be at the time looking at the heavens, by the increase of general illumination of the atmosphere, resembling the effect of young Moon*. The source of von Humboldt was a note on α Centaury written by W. S. Jacob, an engineer and amateur astronomer that was working in India. In 1848 Jacob sent that note to his friend Piazzi Smyth, who presented the work to the Royal Society and then published it (Piazzi Smyth 1849); in that year Jacob was appointed director of the astronomical observatory of Madras. In the note it is possible to read also that the *excessive splendour is caused not only by the profusion of first, second, and third magnitude stars in the neighbourhood, but by the extraordinary general breadth and brightness of the Milky Way thereabouts*. Note that 170 years ago the real structure of the Milky Way was not known. Jacob thought he was looking at the southern part of the vast ring of the Milky Way surrounding us, and he writes that *the superior brightness of so a large proportion of the stars is then naturally* *accounted for by the greater proximity to us*. That is, we were not in the centre of the ring of the Milky Way, but closer to its southern part.

This is the testimony of astronomers of two centuries ago about a spectacular light effect. Two centuries ago the night sky probably was still as dark as many centuries and millennia before. For example, the street lightning in the towns for most part of the eighteenth century was with gas lamp and not yet with electricity; at Bombay in India such gas lamps did not exist until about 1863. Therefore we think that the "Jacob effect" could have some importance for archaeoastronomy, since such a spectacular sky had been visible in the Mediterranean basin until the end of the Bronze Age before disappearing owing to the precession. That is, during the long nights in wintertime, when the Moon was lacking or not yet risen, the ancestors could enjoy for many hours a sort of twilight.

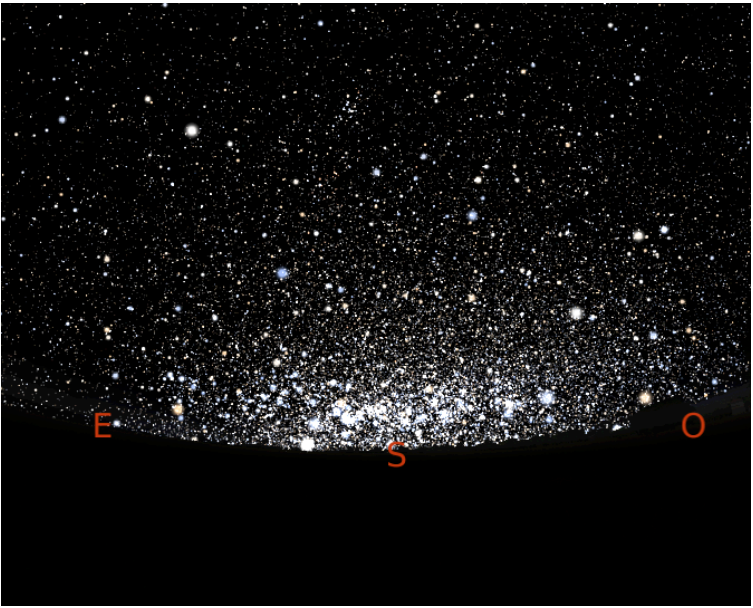

**Fig. 5.** A simulation of the sky visible in Palestine during biblical times (1st millennium BCE; program STELLARIUM). The brightest stars near the horizon in the South-East direction are α and β Centauri and those of the Crux.

## 4. Simulations

The question now is how to deal with this issue. Has one to go to lonely southern places with the darkest sky in order to verify it? Today, in the best observing sites, that is those with the largest telescopes and the darkest skies, the astronomers live in rooms strongly illuminated and it takes a lot of time to accustom the eyes to dark. To spend that time outside the buildings is rather uncomfortable in such places, owing to cool and often windy weather. Therefore to verify such an effect today is not an easy job. Moreover, subjective opinions should be avoided, and therefore we think that the only reliable attempt would be to make accurate simulations.

The simulations require the use of the modern star catalogues containing many million stars, and the application of some astrophysics, since we have to take into account the Rayleigh scattering in the atmosphere and other physical effects. This is a long term project, and we show here just one simple example (Figure 6). We have projected TYCHO2 (Hog et al. 2000) star brightness using IDL. It is possible to see that the brightest part of the Milky Way is located indeed at southern declinations, near the Southern Cross. When we include about 100 million stars of the UCAC3 catalogue (Zacharias et al. 2010; $V < 16$), the brightness of the Sagittarius region is similar to that of the Crux region, and one would expect an analogous Jacob effect even in this case. Since this second case was not mentioned, a study is needed that take carefully into account the geographical position of the observer, the date and time.

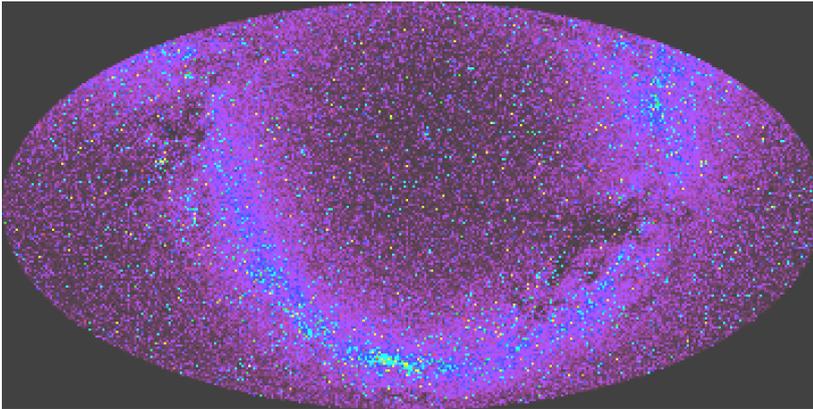

**Fig. 6**. Mollweide (equal area) projection of the brightness of about 2.5 million stars in the TYCHO 2 catalogue (V < 12 mag). Abscissae: right ascension; ordinatae: declination. Each point is one square degree. The brightest part of the Milky Way is that of the Centaurus – Crux region (lower part of the figure).

Our preliminary conclusion is a warning for archaeoastronomers. One has to be careful before claiming that our ancestors looked at a specific celestial object. In our case, it could be possible that the ancestors, even though they used a stellar target for the orientations, were not interested in the bright star itself.